\documentclass[sn-apa,Numbered]{sn-jnl}
\usepackage[T2A,T1]{fontenc}
\usepackage{CJKutf8}

\usepackage{times}
\usepackage{graphicx}%
\usepackage{multirow}%
\usepackage{amsmath,amssymb,amsfonts}%
\usepackage{amsthm}%
\usepackage{mathrsfs}%
\usepackage[title]{appendix}%
\usepackage{xcolor}%
\usepackage{textcomp}%
\usepackage{manyfoot}%
\usepackage{booktabs}%
\usepackage{algorithm}%
\usepackage{algorithmicx}%
\usepackage{algpseudocode}%
\usepackage{listings}%

\begin{document}

\title[Diversity and Language Technology]{Diversity and Language Technology: How~Techno-Linguistic~Bias Can~Cause~Epistemic~Injustice}

\author{Paula Helm\footnote{University of Amsterdam, The Netherlands.}, Gábor Bella\footnote{University of Trento, Italy; IMT Atlantique, France.},  Gertraud Koch\footnote{University of Hamburg, Germany.}, and Fausto Giunchiglia\footnote{University of Trento, Italy.}}
\date{}




\abstract{
It is well known that AI-based language technology---large language models, machine translation systems, multilingual dictionaries, and corpora---is currently limited to 2 to 3 percent of the world's most widely spoken and/or financially and politically best supported languages. In response, recent research efforts have sought to extend the reach of AI technology to ``underserved languages.'' In this paper, we show that many of these attempts produce flawed solutions that adhere to a hard-wired representational preference for certain languages, which we call techno-linguistic bias. Techno-linguistic bias is distinct from the well-established phenomenon of linguistic bias as it does not concern the languages represented but rather the design of the technologies. As we show through the paper, techno-linguistic bias can result in systems that can only express concepts that are part of the language and culture of dominant powers, unable to correctly represent concepts from other communities. We argue that at the root of this problem lies a systematic tendency of technology developer communities to apply a simplistic understanding of diversity which does not do justice to the more profound differences that languages, and ultimately the communities that speak them, embody. Drawing on the concept of epistemic injustice, we point to the broader sociopolitical consequences of the bias we identify and show how it can lead not only to a disregard for valuable aspects of diversity but also to an under-representation of the needs and diverse worldviews of marginalized language communities.}

\maketitle

\section{Introduction}

At the latest since the release of products such as DeepL or ChatGPT, AI-supported language technologies are well on their way to becoming mainstream and thus an integral part of everyday communication and work routines. As such, they shape social relationships and influence processes of knowledge production and proliferation. Following science and technology scholar Langdon Winner, language technologies can be defined as inherently political because they warrant processes of profound social change \citep{Winner_1988}. Given that language technologies are both sociotechnical and inherently political, the question is how they privilege certain points of view and how their specific design is influenced by the interests and idea(l)s of certain groups of people. From an ethical point of view, LT applications thus require appropriate reflection of their inherent biases to prevent discriminatory consequences for marginalized groups of people.

When it comes to questions of linguistic bias in NLP, \cite{Hovy_Prabhumoye_2021} identify five sources of biases: 1) the data, 2) the annotation process, 3) the models, 4) the input representation process and 5) the research design for studying the biases. Following up on this, \cite{blodgett2020language} develop three recommendations for further research: a.) better consideration of social hierarchies and the language ideologies created or transported through the systems, b.) the specification of the normative dimensions applied in the analysis with respect to what is harmful and what beneficial when systems are applied, and c) value-sensitive and community oriented perspectives of  NLP systems in use in language practice. While we take these recommendations as a starting point, we also acknowledge two limitations: one is the focus on linguistic dimensions, which does not pay sufficient attention to the biases in the design of the respective technologies and the methodologies behind them, and the other is the lack of a more explicit discussion of the ethico-political dimensions of the problem.

With this paper, we aim to take a first step towards filling these two gaps by extending the understanding of bias in NLP to another dimension of bias at the technological level: hard-wired and but mostly unintentional by-design preferences for certain languages. We further point out the discriminatory consequences caused by what we call \textit{techno-linguistic bias}. These are, we admit, difficult to pin down, perhaps even more so than is the case with racial or sexist biases in NLP \citep{Bender_Gebru_McMillan-Major_Shmitchell_2021}. This is because the type of bias we are pointing at unfolds its problematic effects not primarily at the individual but at the systemic level. Nevertheless, it is important to pay attention to it because it has far-reaching effects on the equal opportunities of different language communities in terms of self-representation, epistemic self-determination, and communicative participation \citep{Nyabola_2018}. 

Apart from the research around linguistic bias in NLP, recent debates on what has been termed the \emph{digital language divide} are also central when dealing with techno-linguistic bias \citep{Zaugg_Hossain_Molloy_2022, Young_2015}. Digital language divide has been coined to refer to the gap between languages with and without a considerable representation within the worldwide digital infrastructure. As shown by \cite{kornai2013digital} about 10~years ago, less than~5\% of the world's 7-8,000 languages have a remotely significant representation on the Internet, and despite the progresses of a decade, the gap has barely shrunk. The political dimension of the divide is most evident when reconstructing the argument of size, which of course matters in the rapid upscaling of digital support for certain languages, but is at once a result of imperialist politics and far from the only determining factor in digital support. Consider, for example, that the number of Wikipedia pages for Kiswahili, one of the major African languages with about 80~million speakers, is as high as for Breton, an endangered Celtic language in western France with about 200~thousand speakers (according to optimistic estimates). The former, although widely spoken receives little support and if so, mostly top-down, while the latter benefits from generous culture preservation programs.

For many members of language communities such as Kiswahili, digital representation is an urgent project \citep{Nyabola_2018}. Following this demand, in the field of language technology, riding the wave of the recent breakthrough of neural AI, the last decade saw a surge in multilingual language tools and resources for `under-resourced languages.' Researchers have tried to enable technologies such as machine translation, natural language processing, or speech recognition, to an ever larger scope of languages. However, many such efforts are based on a simplistic vision according to which, with the help of AI, already successfully developed and applied methods and systems that are designed and sought of from an anglocentric culture of technology development, are one-to-one adopted to other contexts \citep{bird-2020-decolonising,schwartz-2022-primum}. This simplistic approach to bridging the divide leads to a misalignment between the interests and solutions of the former and the living realities of the latter \citep{Helm_Goetzen_Cernuzzi_Diwakar_Hume_Ruiz-Correa_Gommesen_Gatica-Perez_2022}. Worse, due to general ignorance of the more profound dimensions of linguistic diversity and ultimately the cultural differences that this diversity embodies, major quality problems in the results are neglected, which can eventually result in subtle but nonetheless far-reaching forms of westernized cultural homogenization and epistemic injustice \citep{Spivak_1988}. 

\textit{For these reasons, in this paper, we do not echo the call for bridging the divide by simply digitizing and integrating all the world's languages into existing large-scale technological infrastructures. Instead, what we are concerned with, is the structural inequalities expressed in the disparity of linguistic representations, the causes and consequences of that inequality, and the question how it can be addressed in ways that do not end up contributing to neocolonial dynamics.} The first of our paper's three contributions is to define and outline in more detail the phenomenon of techno-linguistic bias. A resource or tool exhibits techno-linguistic bias if, by design, it is unable to adequately represent or process the language to which it is applied. As we show with several cases of lexical gaps, techno-linguistic bias is closely related to a second key concept of our paper, techno-linguistic diversity, which refers to linguistic constructs and ultimately ideas that are difficult to translate into other languages. We argue that if we are to do justice to the more profound dimensions of epistemic diversity expressed in different languages and prevent epistemic injustice by means of technological expansion, we need to pay attention to precisely these constructs and the cultural particularities they reveal.

In our second contribution, we detail how the causes of techno-linguistic bias are in part a consequence of the flawed methods by which technology is currently developed. In doing so, we refer primarily to the academic apparatus of knowledge and technology production, but also outline how this apparatus is intertwined with the private sector and its interests. To support our claim, we analyze how many databases and language processing systems that are purportedly multilingual have been developed from the perspective of a single language (primarily English). Without a comprehensive understanding of linguistic diversity that goes beyond simply representing another language in a pre-existing model, they run the risk of developing a hard-wired bias that may superficially serve to fill a language gap, but on closer inspection is ultimately ineffective at supporting the values that closing the gap is intended to promote.

Our third contribution addresses the consequences of efforts to represent diversity in a simplistic way, such as the representation of lexical and semantic data, without considering the broader implications. We show how simplistic representations of diversity can lead to inevitably false representations of particular languages, which, when they penetrate previously under-served communities, can lead to dialectic dynamics and support existing or new forms of epistemic injustice, which we outline below in more detail. As an alternative, we make a case on the grounds of co-design that reflects, supports and accounts for diversity in a much more principled and systemic manner than any top-down approach can \citep{Saad-Sulonen_Eriksson_Halskov_Karasti_Vines_2018, Smith_2021}. 

In line with these contributions, the rest of the paper is organized as follows. In Section~2, we define and discuss the notion of (techno-linguistic) diversity as well as epistemic injustice, respectively. Section~3 is devoted to the definition of (techno-linguistic) bias,  the various forms it can take and its causes. Section~4 tackles the normative consequences that follow from the techno-linguistic bias we identified. Finally, in Section~5, we discuss approaches to co- and participatory design in language technology and clarify some of the conditions that we see need to be fulfilled in order to avoid the ethical dangers identified. 

\section{Diversity as an Ethical Norm}

Although our own point of departure is the normative one of protecting, promoting, and preserving diversity, we are wary of the problems that come with naively celebrating it without proper conceptualization \citep{Helm_Michael_Schelenz_2022}. Acknowledging that diversity is a moral-epistemic hybrid \citep{Potthast_2014}, we differentiate between an understanding of linguistic diversity as a \textit{descriptive} and a \textit{normative} concept, to  better distinguish between (a)~the actual notions of difference that underlie our understanding of linguistic diversity as a design strategy, and (b)~the values we associate with diversity as the objective of our work. 

\subsection{Diversity: a Moral--Epsitemic Hybrid}  
When reviewing the ethics policies of large tech companies, diversity is regularly listed as one of the core corporate values. However, a closer look reveals that the presentation in such guidelines lacks complexity. Instead, diversity is often mentioned only superficially and reduced to simplistic but easily measured and scalable categories such as gender, race, or age, resulting in what Benjamin has so aptly observed as ``cosmetic diversity'', which ``too easily stands in for substantive change'' \citep[p.~24]{Benjamin_2019}. Cosmetic diversity is problematic for several reasons \citep{Chi.2021}. First, because it clouds our eyes to the ambiguity of diversity as an instrumental and thus conditional value. Second, such portrayals, which have become part of the rhetoric of many large technology companies in response to various scandals, often view and treat diversity as a kind of resource that can be ``exploited.'' Political philosopher Iris Young, however, warned already in the 1980s against such capitalist appropriations of the concept, where diversity is instrumentalized as something that ``enriches me'' or as a means of optimally valorizing people or enhancing the performance of institutions and organizations. Instead, diversity is about how we can live together in pluralistic societies in an inclusive, participatory, and nondiscriminatory way \citep{Young_1990}. 

To clarify this difference, anthropologist Anna Tsing speaks of ``meaningful diversity", that is, diversity that changes things,' as opposed to scalable diversity, which accepts only what can be incorporated into pre-existing standards without further adaptation \citep{Tsing_2012}. Tsings distinction between meaningful and scalable diversity is instructive here, as it highlights exactly the difference that we want to point out when criticising current attempts to increase linguistic diversity in language technology, in ways that simply extend systems already in place. These attempts, we will show, fail to account for the more profound cultural and epistemological differences, which are incorporated within different languages and which, as we claim, should be at the heart of diversification efforts. This, however, would require much more profound adoptions all the way through the methodological, design, engineering and implementation circle.
 
Understood in the ways proposed by Young and Tsing, diversity is a contested notion that is not always to be embraced unconditionally. Yet, it is decisive for broader value based projects such as decolonization \citep{Vertovec.2012}. The UNESCO Convention on the Protection and Promotion of the Diversity of Cultural Expressions supports this idea of linking the preservation of diversity with values such as tolerance, inclusion, and dignity \citep{UNESCO.2005} However, it is also important to recognize the limits of the value of diversity. Whether diversity actually helps to support broader goals, crucially depends on the ways we understand it, and following from that, what we are willing to do to put it into practice accordingly. 

\subsection{Linguistic Diversity}

As a design strategy, then again, diversity helps define differences between entities, such as languages, and point out their unique features (e.g.~words or expressions that cannot be translated easily into other languages, notions that only make sense to specific speaker communities). The terms \emph{language diversity} and \emph{linguistic diversity} are often used to refer to the over seven thousand languages existing in the world, and to the wide-ranging differences among them \citep{giunchiglia2018one}. The association of diversity to language implies the preservation of the variedness of the world's linguistic landscape. In the field of linguistics, diversity is not a technical term and is therefore usually used in an informal way, with a few notable exceptions. \cite{greenberg1956measurement} defined linguistic diversity as the probability of two persons speaking the same language in a certain geographic area. \cite{rijkhoff1993method}, instead, apply the term (informally) to sets of languages, and understands the `variedness' of the languages in terms of their genetic relationships.

The following definition of techno-linguistic diversity, in its very specific perspective will help us to critically scrutinize existing attempts to close the digital language divide toward their ability to actually represent differences between languages well enough to do justice to the normative dimensions of diversity:

\vspace{2mm}
\begin{quotation}
\noindent\textit{Techno-linguistic diversity applies when a language technology is able to process and represent different linguistic means available in different languages to express ideas, objects, worldviews, relationships, or observations, even when the most well represented languages do not provide an equivalent means and thus can only indirectly or approximately express the idea. }
\end{quotation}
\vspace{2mm}

The most straightforward examples of linguistic diversity are found in lexical semantics, in relation to the well-known phenomenon of untranslatability. One example from the domain of kinship (the diversity of which is well documented) is the Maori word \emph{teina}: it means \emph{elder brother} if it is pronounced by a male speaker, and \emph{elder sister} if it is pronounced by a female. In translation to English, this concept can only be expressed in an approximate way. Another example is the phenomenon of \emph{inalienable possession}, widely present in Native American and Australasian languages, where abstract---yet for us natural---concepts such as \emph{mother} or \emph{head} (as a body part) cannot be expressed as single words (free morphemes), but only together with their possessor (i.e.~as the combination of two bound morphemes): \emph{my mother}, \emph{your head}.

Motivating our normative stance on the importance of properly dealing with diversity when building or expanding  language technology, we claim that, for native speakers, such language-specific terms are often inextricably embedded in the local context. For a speaker in South India, choosing the correct term out of 16~possible terms to designate one's cousin---depending on gender, age, the mother's or father's side, etc.---is a basic requirement of politeness and culture, while in other languages, there is only one single term existing for cousin. Although kinship is a prime example of linguistic diversity, it can also be reflecting of geographical specifics of particular regions. For example, in the Italian Alps, the word \emph{malga}, designating a typical mountain restaurant with no equivalent outside the Alpine region, is an important everyday term with a strong connection to south alpine tradition and culture.

From the perspective of computational linguists and engineers, in contrast, diversity represents a boundary beyond which algorithms do not scale. Given the persistent and increasing scaling pressures in the field, which we will outline in more detail in the next section, it is a well-understood temptation to simply ignore such long-tail phenomena and concentrate on the more high-level representation where diversity at scale is relatively easily achieved.  
Yet, it is not always impossible to reconcile the engineer's inclination for automation with an accurate computational representation of linguistic diversity. One solution is to rely on the vast scientific data on linguistic typology produced by experts through the last century.  \cite{giunchiglia2017understanding} used a quantified measure of the diversity of sets of languages for the prediction of the universality or specificity of linguistic phenomena. \cite{khishigsuren2022using} used results from in-depth, local field studies to better understand the meaning of family relations in order to produce accurate kinship terminologies in no less than 600 languages. In \cite{bella2020major}, an about 10-thousand-word formal lexicon of Scottish Gaelic was co-created by local language experts, including locally specific terms not directly translatable to English or most other languages. 

These examples show that the representation of linguistic diversity in language technology is not only a question of feasibility, but also a question of normative orientation and the related priorizations leading to an intensified investment in engagements with local communities and co-creation efforts.

\subsection{Epistemic Injustice}

We have already pointed out the importance of rigorous conceptual work for the meaning we attach to the normative concepts that guide our efforts. It is equally important to clarify what is lost or which kinds of harms are done when these norms are violated, that is, when diversity is simplified in such a way that its normative dimensions are eroded. In the introduction, we repeatedly used the term ``epistemic injustice'' because it not only describes well the homogenization that can result from loss of diversity, but also situates that loss and the attached harms within a broader context of global inequalities.

The term ''epistemic injustice'' was introduced by philosopher Miranda \cite{Fricker_2009} and refers to a typology of injustice that is distinct from the injustice caused by the inequitable distribution of epistemic goods, such as educational materials, books, or information technologies. It is therefore very useful in accurately understanding and naming the problems that arise when techno-linguistic bias persists despite, or because of, the broad extension of language technologies to a variety of languages and communities. Rather than focusing on the issue of distribution of resources  \citep{Goldman_2002}, epistemic injustice, as understood here, addresses the harms that occur at a more subtle level when people are unequally valued in their capacity as bearers and practitioners of different forms of knowledge \citep{coady_2010}. According to Fricker's analysis, the most important forms of epistemic injustice include forms of exclusion and silencing, the systematic distortion or misrepresentation of certain people's meanings or contributions, and the undervaluing of their status in communicative practices. 

Epistemic injustice also has a clear political connotation in that it disproportionately affects groups of people who are already disadvantaged because of their social identities, such as race, gender, class, or disability. In addition to the inequitable distribution of resources, epistemic injustice affects the ways in which knowledge and experiences are recognized, valued, or discredited by others. It manifests itself in two main forms: testimonial and hermeneutic injustice. The second of which is most relevant to the present case. Hermeneutic injustice refers to a situation in which a person or group is disadvantaged because their experiences or social realities are not acknowledged or understood due to a lack of concepts, vocabulary, or frameworks. In such cases, it may be difficult for individuals to articulate their experiences or seek redress because this particular, rather subtle but no less relevant form of injustice is not adequately recognized or understood by society \citep{Fricker_2009}.

From an overarching perspective, the concept of epistemic injustice also needs to be situated historically, as it can be understood as a further development of Gayatri Chakravorty Spivak's notion of epistemic colonization \citep{Spivak_1988}. Epistemic colonization refers to the processes by which one's culture's knowledge systems, beliefs, and ways of knowing are imposed on another culture or community, often as a direct, indirect, or late consequence of colonization or imperialism. This involves the domination of a particular theory of knowledge (in the present case, it may be a belief in the universal power of AI systems developed in the West) over others, often marginalizing or suppressing local knowledge systems and ways of understanding the world.

Epistemic injustice, understood as rooted in a history of epistemic colonization, can lead not only to individual but also to structural harm, as it is usually accompanied by a loss of cultural diversity and leads to a form of homogenization or violent cultural appropriation that ultimately benefits those who caused the injustice. In this way, existing power imbalances are perpetuated as imperialist knowledge becomes the standard against which all other knowledge is measured. This can entrench structural dependencies. Epistemic injustice, as we understand it here and use it to critically assess the effects caused by current initiatives to expand language technology, builds on historically established inequalities and need to be understood in this cronyism. In our view, counter-designs and strategies can only function if they take this broader context into account. 

In what follows, we lay out how recent attempts to close the language gap through distributing epistemic resources but without accounting for meaningful diversity are at risk of contributing to epistemic injustice. To do so, we elaborate on what techno-linguistic bias means as a counterpart to techno-linguistic diversity, and why it is a much more political matter than it might at first appear from a purely technological or linguistic perspective.

\section{Techno-Linguistic Bias in Language Technology}

In the context of AI language technology, the notion of \textit{bias} has so far been used to refer to patterns of stereotypes and preferences towards social groups, most often concerning learning-based language processing systems \citep{blodgett2020language}. In terms of social groups, studies have mostly focused on gender, ethnicity, and race, but also other forms of bias (religion-related, age-related, political, etc.) \citep{FriedmanNissenbaum1996}. Bias has gained much attention and was prominently problematized as it has been identified as one of various sources of automated discrimination \citep{Barocas_Selbst_2016}. 

While we agree that it is important to problematize bias, we also recognise that bias is omnipresent and that, even if it is usually associated with a negative connotation, it actually need not be harmful to diversity \textit{per se}. For example, when affirmative action serves to counteract the unequal representation of otherwise marginalized groups, bias may well be intentional and desirable. Contrary to a blanket critique of bias as a phenomenon in itself, we accept that all knowledge, all insights, and even all data are situated, i.e.~they always reflect a particular point of view in space and time that is influenced by culture, history, politics, economics, epistemology, and so on \citep{Haraway.1988, Gitelman_2013}. 

Unbiasedness is therefore a deceptive goal that, instead of solving social problems, reproduces problematic ideas, such as the unrealistic imaginary that technology can be neutral \citep{Beer_2017}. It is therefore important to be upfront about when and for what reasons a certain bias is problematic and needs to be combated, and that this combating does usually not lead to no bias, but to a different, ideally more just bias \citep{Harding_1995}. Bias is harmful to diversity when it leads to the misrepresentation of already vulnerable, underrepresented, and marginalized groups. Such bias calls for counteraction. Similarly, when large-scale language technologies are biased towards the correct representation of languages of colonial powers, but disregard the particularities of other languages that are also spoken by many people or are at risk of extinction, this demands change. To enact such change sustainably, it is not just instrumental to develop strategies to counteract techno-linguistic biases, but also to invest work into unraveling their causes.

\subsection{Techno-Linguistic Bias}

Our focus is a specific, so far not widely studied form of bias that we call \emph{techno-linguistic bias}. Techno-linguistic bias goes beyond linguistic biases as known in communication research. The subject of techno-linguistic bias are not just languages \textit{per se} but includes the design of language technology: corpora, lexical databases, dictionaries, machine translation systems, word vector models, etc. Techno-linguitsic bias is present in all of them, but it is easiest to observe with respect to multilingual resources and tools, where the relative correctness and completeness for each language can be observed and compared. We define it as follows:

\vspace{2mm}
\begin{quotation}
\noindent\textit{Techno-linguistic bias is observed when the technology, by design, represents, interprets, or processes content less accurately in certain languages than in others, thereby forcing speakers of the disadvantaged language to simplify or adapt their communication, (self-)representation, and expression when using that technology to fit the default incorporated in the privileged language.}
\end{quotation}
\vspace{2mm}

Bias manifests itself through linguistic or cultural inaccuracies in the way a language is processed or represented. By emphasising the \emph{by-design} aspect of techno-linguistic bias, our definition is deliberately focusing on the representational, rather than the allocational harm of bias in language technology. Thus, techno-linguistic bias is not primarily concerned with the scarcity of data itself or with biases within language, but rather with structural bias built into language processing algorithms, representational models, resources, or methodologies.

The social groups affected by techno-linguistic bias are clearly the communities of speakers of underrepresented languages, however heterogeneous they may be otherwise (according to social status, culture, gender, race, ethnicity, religion, etc.).  Being the native or second-language speaker of a language determines one's access to information, and the language technology that enables this access affects one's ability to communicate, on the Web or elsewhere. To our knowledge, the term \emph{techno-linguistic bias} has not been used as a analytical device or design strategy in any way similar to ours while many of the underlying neocolonial mechanisms have, however, been pointed out  \citep{bird-2022-local, schwartz-2022-primum} .
In terms of actual bias in AI systems and data, research concerning the representation (or the lack thereof) of the vernaculars of social groups within language resources is closest to ours.  Here, however, we want to go a step further in pointing out that in recent attempts, both in the field of engineering and technology advancement and in the fields of ethics, policy and development aid, the language communities themselves are left out of the process. These attempts or projects, with \citep{Aradau_Blanke_2022} can be described as techniques of governing emerging technology, which while striving for diversity as one of their goals, turn those most affected by the results into what philosopher Jacques Rancière has called the ``part of those who have no part'' \citep[p. 30]{Ranciere_1998}. It is the technology developers and designers residing in the big companies as well as influential academic institutions that fashion themselves as the experts who are called upon to embed linguistic diversity into their tools and expand them under the normative guise of inclusion. However, in the process, techno-linguistic bias is reproduced because the Western perspective is taken as the norm and the subjects of diversity remain at the outside. It is this form of exclusion which at the same time allows for and is mobilized by investments in scalable diversity.   

\subsection{Forms of Techno-Linguistic Bias}
\label{sec:bias_technology}

In order to illustrate what we are talking about more concretely, in the following we provide examples of techno-linguistic bias from two kinds of language technologies widely used today: multilingual databases, and machine translation systems.

\begin{figure}
    \centering
    \includegraphics[width=.8\textwidth]{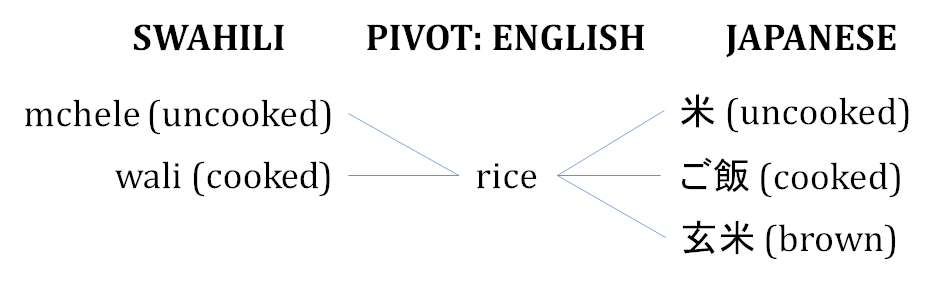}
    \caption{Biased cross-lingual mapping of words about various forms of `rice' from a popular multilingual lexical database.}
    \label{fig:mldb_mapping}
\end{figure}

\begin{figure*}[t]
    \centering
    \includegraphics[width=.9\textwidth]{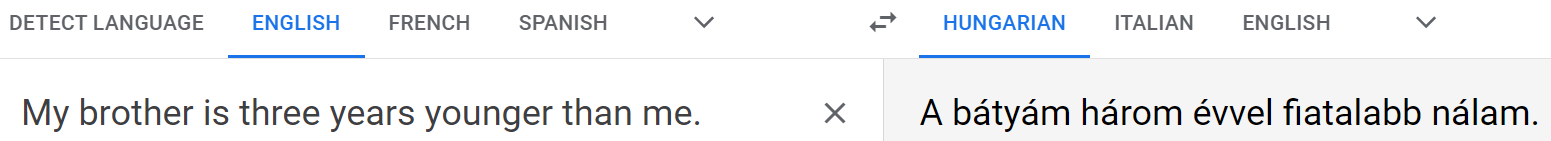}\\
    (a)\\
    \includegraphics[width=.9\textwidth]{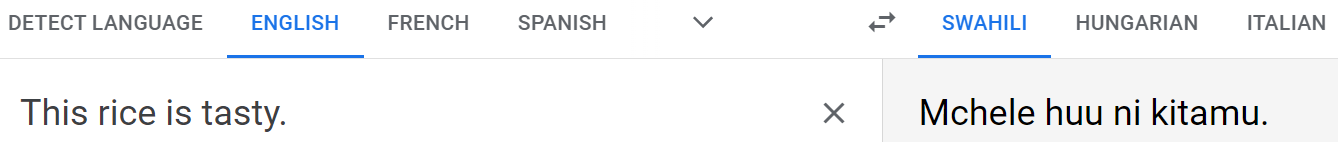}\\
    (b)\\
    \includegraphics[width=.9\textwidth]{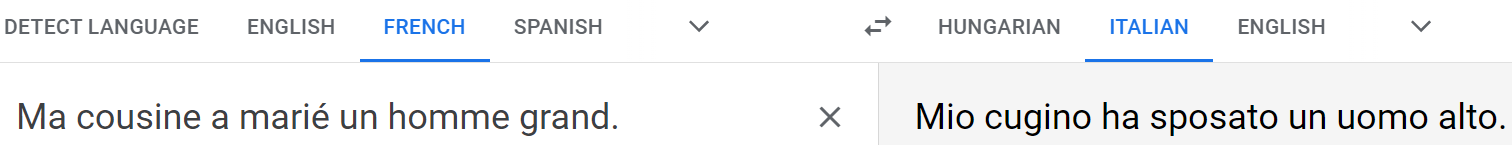}\\
    (c)\\
\caption{Examples of techno-linguistic bias in machine translation. (a)~The lack of an equivalent common word in Hungarian for \emph{brother} results in an erroneous translation meaning \emph{my elder brother is three years younger than me.} (b)~The lack of an equivalent term for rice in Swahili results in an erroneous translation meaning \emph{this raw rice is tasty.} (c)~The systematic use of English as pivot language results in an erroneous change of gender when translating between French and Italian.}
    \label{fig:mt}
\end{figure*}

As a generalisation of bilingual dictionaries, the 2000s saw the appearance of \emph{multilingual lexical databases} that map words, based on their meanings, across large numbers of languages. While these resources proved to be extremely useful in cross-lingual applications, looking under the hood---into their underlying models of lexical meaning---reveals varying levels of limited expressivity and bias.

As shown by \cite{giunchiglia2023representing}, several of these multilingual databases interconnect words from hundreds of languages, mapping the words of each language to 100~thousand English word meanings (so-called \emph{synsets}) of Princeton WordNet \citep{miller1998wordnet}. On the one hand, this choice makes practical sense, as among all similar resources, WordNet offers by far the widest coverage of word meanings. On the other hand, this results in a strong bias towards the English language and Anglo-Saxon culture in general, as the expressivity of the database is limited to notions for which a word exists in English \citep{giunchiglia2023representing,bella2022linguistic}. Figure~\ref{fig:mldb_mapping} provides a simple example from the food domain, known to be culturally, and thus also linguistically, diverse. It shows how a biased lexical database maps together words in Swahili and Japanese meaning \emph{uncooked rice}, \emph{cooked rice}, and \emph{brown rice}. The degree of information loss is flagrant: while both Swahili and Japanese provide fine-grained lexicalizations about the various forms of rice, the many-to-many mapping that results from passing through English masks all fine-grained differences, resulting in both a loss of detail and incorrect translations when one moves from Swahili to Japanese or vice versa. The diversity-diminishing bias towards the English language and Anglo-Saxon cultures is also found in other domains that are well-known to be diverse across languages: family relationships, school systems, etc.

Not all multilingual lexical resources are affected by Anglo-Saxon bias. The linguist community, and in particular typologists, have been studying similar phenomena for a long time. The results of this research, however, are primarily destined to researchers in historical and comparative linguistics and are not directly usable outside these specialist communities.


\textit{Machine Translation}
(MT) is the flagship task of AI-based language technology. Without claiming to be exhaustive, we point out three aspects of current MT technologies where techno-lingusitic bias can be observed: the non-handling of untranslatability, the variedness of vocabulary and grammar, and the use of a pivot language.

Today's top MT systems, such as DeepL and Google Translate, make systematic mistakes over untranslatable terms, betraying the fact that this phenomenon is not specifically addressed by these tools \citep{khishigsuren2022using}. The screenshots~(a) and~(b) in Figure~\ref{fig:mt}, taken from a mainstream machine translator, show examples of erroneous translations due to untranslatability. For instance, when translating the English sentence \textit{My brother is three years younger than me} to Hungarian, Mongolian, Korean, or Japanese, syntactically correct yet semantically absurd results are obtained:

\vspace{2mm}
\begin{center}
    Hungarian: \textit{A bátyám három évvel fiatalabb nálam.}\\
    Japanese: \begin{CJK}{UTF8}{min}私の兄は私より3歳年下ですす.\end{CJK}\\
    Korean: \begin{CJK}{UTF8}{mj}형은 나보다 세 살 아래다.\end{CJK}\\
    Mongolian: {\fontencoding{T2A}\selectfont Ах маань надаас гурван насаар дүү.}
\end{center}
\vspace{2mm}

These languages either have no equivalent word for \emph{brother} (as in Mongolian) or, when they do, the equivalent word is rare (as \emph{fiútestvér} in Hungarian). Based on training corpus frequencies, the MT system ends up choosing a semantically unsuitable word, such as \emph{bátyám} meaning \emph{my elder brother}, resulting in \emph{My elder brother is three years younger than me.} A similar example, based on the example of \emph{rice} from earlier in this section, is the following English-to-Swahili translation:

\vspace{2mm}
\begin{center}
    English: \emph{This rice is tasty.}\\
    Swahili: \emph{Mchele huu ni kitamu.}
\end{center}
\vspace{2mm}
\noindent where the Swhaili sentence means \emph{this raw rice is tasty}. Note that these are not cherry-picked exceptions but rather are examples of systematic mistakes within domains of high linguistic diversity.

A second form of bias concerns the variedness of vocabulary and grammar in MT output.  \cite{vanmassenhove-etal-2021-machine}  quantitatively compare the lexical and grammatical diversity between original and machine-translated text. Their definitions of diversity and bias are different from ours: by diversity they refer to the richness of the vocabulary and the complexity of the grammar of a document (normalised by document size and computed according to various metrics), while by bias they understand an uncontrolled loss of diversity due to MT. Still, their results are also relevant under our interpretation: They reports that, for the same language, morphology in translated text becomes poorer with respect to original (untranslated) corpora, i.e.~features of number or gender for nouns tend to decrease. This phenomenon affects morphologically rich languages in particular. 

A third form of techno-linguistic bias in MT is their use of English as a pivot language when translating between non-English language pairs. This practice is explained by the relative scarcity of bilingual training corpora for such language pairs, as well as scalability. Example~(c) in Figure~\ref{fig:mt} shows the case of French-to-Italian translation of a sentence meaning \emph{my (female) cousin married a tall man.} While French and Italian (as do most languages) use different words for male and female cousins (\emph{cousin/cousine}, \emph{cugino/cugina}), English does not. The result is that the gender of the cousin is `lost in translation' and, as a form of combined linguistic and gender bias, it appears as a male in the translation.

\subsection{Causes of Techno-Linguistic Bias}
\label{sec:bias_methodology}
Because bias is most problematic when it reflects and perpetuates existing power relations, any critique of technological bias should ideally include at least a brief genealogy of the origins of bias and the ways in which different social groups are harmed or benefited in different ways. The following subsection highlights how well-intentioned but unreflective attitudes in computational linguistics research and development practices contribute to the creation of language technologies that are adverse to meaningful diversity. It also reveals that computational linguistics has not developed in isolation, but is situated within and reflects broader global power dynamics. Reflecting on the situatedness of this kind of research is useful for clarifying why focusing on language inclusion and expansion is not enough to promote diversity, but may ultimately even reproduce entrenched dynamics of epistemic injustice.
In the last 50~years, research in Computer Science has been dominated by a strong Anglo-Saxon influence, reflecting turn-of-century power dynamics. In Computational Linguistics, this bias was apparent across all prestigious publications, conferences, and journals of the field: an unspoken convention required for research to be considered as competent to be applied and demonstrated in English. Thus, English has not only been the \textit{lingua franca} of scientific communication, but also the \textit{de facto} standard subject matter of research. This is not to say that scientific results on other languages were not published, but they were  generally considered by the community (paper reviewers, journal editors, etc.) as `language-specific' and therefore less likely to be relevant to a wide audience. Publications about languages other than English were relegated to second-tier or niche journals and venues. \cite{schwartz-2022-primum} reports that between 2013 and 2021, 83\% of papers accepted at ACL---the flagship conference in Computational Linguistics---were explicitly or implicitly about English and 97\% were about Indo-European languages. 

Despite these numbers, the 2010s saw an emerging interest in multilingual language technology, and of a new research sub-field targeting `low-resource' (or `under-resourced') languages, previously neglected by mainstream research. This change of scope is tightly related to the blazing progress of deep-learning-based AI on English (and also on some other well-supported languages such as Spanish or Chinese). Problems that were earlier considered as exceedingly hard, such as machine translation, have suddenly been solved with impressive results. For researchers, the `major' languages were not providing suitably interesting challenges anymore, apart from incremental research pushing the accuracy boundary. Low-resource languages seemed like a promising horizon. 

The new fascination with `low-resource languages' does not mean that, say, Mongolian speech synthesis has suddenly become of mainstream scientific interest. In line with the `zero-shot' data-driven ethos \cite{bird-2020-decolonising} of recent deep AI research that shuns any use of prior results from linguists and field workers, low-resource language research is only worthy of a top publication as long as 
(1)~it provides a solution for multiple, preferably tens or hundreds of languages at the same time; 
(2)~it involves mainstream AI technology, i.e.~neural networks; and 
(3)~it requires very little to no knowledge from authors and readers of the languages targeted. 
The typical low-resource research contribution thus scrapes web content, such as Wikipedia pages, written in the languages in question, often without any understanding of their quality or content \cite{lignos-etal-2022-toward}). It then trains or fine-tunes deep learning models based on the data, and finally demonstrates a few percentages of increase in quality (precision, recall, BLEU, etc.) over one of the standard tasks in computational linguistics, such as named entity recognition or machine translation. This practice is certainly not in line with what we earlier described as accounting for meaningful diversity. 

Simultaneously, many highly populated but under-resourced Global South countries were identified by high-tech companies (and digital platforms in particular) as still unsatisfied markets with a potential for data scraping and infrastructural advancements. What happened during the following 10 years has been described as a `race' in which digital platforms swamped African and South East Asian countries, in order to be the first to secure the loyalty of wast new customer bases \citep{Arora_2019, Benjamin_2019}. 

Also, in the field of technology ethics, Silicon Valley has set agendas over the past decade by pumping large amounts of money into an academic system that otherwise faces scarcity measures and budget cuts \citep{Ochigame_2019}. This has led to two types of ethics increasingly taking hold: 1. the type that embraces the notion that it is primarily more technology, and in particular the expansion of AI, whereby existing problems can be solved, and 2. the type of ethics that can be easily transferred into existing systems and infrastructures and from there automated and implemented \textit{en masse}. Both can be demonstrated in the often overly simplistic ways in which racial and gender biases are tackled by developing fairness measures, and can equally be mirrored for the appropriation and handling of calls for better acknowledgment of diversity.

The industrial appropriation of academic ethics research and its influence on the respective notions of justice as fairness and diversity as demographics go hand in hand with the broader ranging imaginary that large scale technological innovation will serve as a panacea for wide-ranging problems \citep{Pfotenhauer_Jasanoff_2017}. Following Anna Tsing, however, we understand scalability not as an intrinsic property of a solution or product (of any kind), but as something that stems from emphasizing certain aspects and omitting others. Ironically, then, for an innovation product to be scalable, it must be designed to reduce the complexity of a problem and its associated solution to isolated parameters that can be fairly easily abstracted from the context of the specific domain or local context for which it was developed \citep{Engel_2016}. This abstraction work makes the innovation generalizable and thus scalable in that the number of languages can be significantly increased without major adaptation \citep{Tsing_2012}. This is exactly what is going on when existing neural language technology is applied indiscriminately to any language without major change or adaptation. 

These problems are exacerbated against the backdrop of a postcolonial computer culture \citep{Irani_Vertesi_Dourish_Philip_Grinter_2010}. In this context, recent public and private sector Data4D efforts have been criticized, not only in terms of their ``white savior'' ethos, but in some cases even to the point of using development goals as free riders to invade vulnerable populations and extract their data \citep{Taylor_Broeders_2015}. Whether intentions are bad or not, the choice of problems to focus on is often driven by either the icentives of the academic communities or by industry pressure or, most likely, a combination thereof. This leads to a gap between the solutions offered and the actual needs of local user communities.

In contrast, language resources that are handcrafted and co-designed, leading to customized solutions to local problems, are difficult to defend as competitive because they are much more complicated to scale as they are, by definition, not generalizable. This is why, if innovation is narrowly defined as the expansion of AI, small languages will continue to be neglected in the field. However, there is a strong case to be made that less high-tech approaches actually offer great potential for social innovation, as they are more likely to address the needs of diverse communities than the top-down, AI and scaling-oriented approaches currently vaunted.

\section{Ethical Concerns with Biases in Language Technology}

The consequences of research being done under these conditions raise a multitude of ethical concerns with regard to potential epistemic injustice being done. Most of these concerns are related to the rather ill-defined attempts to promote diversity by adopting the top-down solutions that AI-approaches warrant and which, by the nature of their design, can only respond to a simplistic idea of diversity. With Western researchers unilaterally setting developmental goals and providing technological solutions to reach them, they effectively and most ironically, silencing the actual speakers. This silencing does not regard the lexical representation or distribution of epistemic goods, which may in fact  be increased, but the types of problem definitions and the corresponding designs of technical solutions \citep{Fricker_2009}. 

Critical commentators on this process such as Bird, Bender et al., or Irani at al. point out that typical research goals such as language documentation and ``technology-based revitalization'' regularly misinterpret the needs of local communities. In most cases, native speakers are not involved in the process, or if they are, they are taking on subordinate roles such as commentator, validator, tester, or worse, data extractor \citep{Helm_Goetzen_Cernuzzi_Diwakar_Hume_Ruiz-Correa_Gommesen_Gatica-Perez_2022}. Instead of co-creating on an equal footing, in many cases the analytical, high-level work is done in technology labs of Western universities or companies, where the languages being studied are sometimes not even understood by the people working on them, let alone the cultures they represent \citep{Arora_2016} . As a result, they are unable to assess the quality of the data on which they rely. Sometimes they do not even know if they are using the right language, as observed in the case of automated Wikipedia scraping \citep{lignos-etal-2022-toward} . 

Added to this is the neglect of the meaning and relevance of language, which goes beyond the mere transmission of information. As a consequence, these dimensions are in danger of getting lost to a culture that limits its normative horizon to technical innovation, linguistic expansion as ends in themselves. However, linguistic research has long recognized that language is more than just a string of words \citep{Bender_Gebru_McMillan-Major_Shmitchell_2021}. The Kenian writer and scholar Ngugi wa \cite{Thiong’o_1986}, for example, has done extensive research on the cultural diversity embedded in Kenya's multilingual heritage. Language conveys the situatedness of knowledge. It influences how we see our world, how we perceive colors and tastes, and even the social and political possibilities of our societies. In Kiswahili, Arabic and Amharic, for example, time is measured not only from one hour to the next, but in relation to sunrise and sunset. Thus, 7 a.m. depends on the season because it is always the first hour after sunrise. This understanding, which better adapts social life to the dynamic rhythm of the year, does not translate one-to-one into the much less dynamic but more definite Anglo-American system for communicating about time, and it would be lost if we all had to squeeze ourselves into English. It is this kind of situated knowledge embedded within languages, which cannot be captured by one-size-fits-all design, nor by AI, no matter how technically sophisticated. It's something that has to be done by people who know what they're talking about, literally.

Given these differences, which go beyond lexical representation, we contend that it is unfair to require all of us to conform to communication norms that have emerged from the perspective of English-speaking (or, increasingly, Chinese-speaking) users if we are to take advantage of language technologies. In other words: can we accept techno-linguistic biases as a fact of the digital age, or do we consider them ethically unacceptable? Following our reasoning, the latter is obviously the case. To go a step further, we argue that the techno-linguistic bias we are currently dealing with are not only passively unfair, but even actively contribute to epistemic injustice. More specifically, it leads to hermeneutic injustice, as we introduce in Section 2.3. Lexical gaps, accordingly, mean in practice that for many people the specifics of their social realities and their resulting situated knowledges are not acknowledged when English is taken as the norm. To defend this rather strong claim, it is important to consider the historically rooted power relations at play here. The political theorist Ali  \cite{Mazrui_Mazrui_1999} has studied the influence of English on African culture and argues that this influence changes the self-perceptions and social practices of African peoples. This need not be bad \textit{per se}. Creole creates something new that we can greatly appreciate; cultural encounters can be enriching and broaden perspectives. The problem, then, is not cultural mixing and matching as such; on the contrary. Rather, it is the dominance of one culture over the others, built on a colonial history of extraction and appropriation, which still exists, albeit in new guises.

\section{Promoting Diversity in Language Technology while~Avoiding Epistemic Injustice}

In our interpretation, accounting for diversity and power asymmetries in language technology is not a fixed state that can be implemented in a system once and for all, but rather a process by which technology is continually adapted to the variety of linguistic phenomena and different communities' needs that it aims to meet. In deliberate opposition to the shortcomings described in Section~\ref{sec:bias_methodology}, \cite{bird-2020-decolonising} and \cite{schwartz-2022-primum} have been advocating alternative, `decolonising' approaches to multilingual research in Computational Linguistics and to working specifically with indigenous linguistic communities. Schwartz considers as prerequisites the obligations of \emph{cognizance} to be aware of the power imbalance, historical traumas, and the difference in epistemology between the researcher and the local community; of \emph{beneficence} to make sure that the research actually benefits the language groups themselves; of \emph{accountability} that involves building working relationships with local people and reporting to them; of \emph{non-maleficence} so that the research does not cause harm. 

We embrace the contributions of Bird, Schwartz, and of design anthropologists such as \cite{Smith_2021} who advocate mutual learning and thus consider all participants in the process simultaneously as researchers and beneficiaries. Specifically with regard to language technology, we again echo Bird's call for a focus on knowledge transfer beyond language, as generational loss of knowledge about local history, practices, etc. is often one of the main reasons for interest in language preservation, and gives rise to deliberate promotion of digitization. Specifically with regard to language technology, we again echo Bird's call for a focus on knowledge transfer beyond language, as generational loss of knowledge about local history, practices, etc. is often one of the main reasons for interest in language preservation, and gives rise to deliberate promotion of digitization.
\begin{table}[t]
    \setlength{\tabcolsep}{2pt}
    \centering
    \begin{tabular}{l|rr|r|rr}
        Language & L1+2 speakers & \% of English & Wiki rank & Articles & \% of English \\\hline
English & 1.45B & 100\% & 1 & 6624314 & 100\% \\
Indonesian & 300M & 20.7\% & 22 & 639717 & 9.7\% \\
Bengali & 300M & 20.7\% & 63 & 134,966 & 2.0\% \\
Marathi & 100M & 6.9\% & 74 & 90,421 & 1.4\% \\
Breton & 200k & 0.01\% & 82 & 78,361 & 1.2\% \\
Swahili & 80M & 5.5\% & 83 & 76,417 & 1.2\% \\
Hausa & 77M & 5.3\% & 123 & 21,190 & 0.3\% \\
Pashto & 40M & 2.8\% & 127 & 17,202 & 0.3\% \\
Scottish Gaelic & 50k & 0.003\% & 133 & 15,859 & 0.2\% \\\hline
    \end{tabular}
    \caption{Contrast of the number of speakers (as first or second language) and the number of Wikipedia articles for a few selected example languages, according to \url{https://meta.wikimedia.org/wiki/List_of_Wikipedias}.}
    \label{tab:languages}
\end{table}

Furthermore, we recognize the  importance of \emph{vehicular} or \emph{trade languages} in addressing local vernaculars. While English is indeed used as \textit{lingua franca} in many parts of the world, it is far from being the only one. Arabic and Persian in the Middle East, Hindi/Urdu in Northern India and Pakistan, Hausa and Swahili in Africa, etc., are also widely used trade languages. Building on this insight, \cite{bird-2022-local} suggests to work with what they call a \emph{multipolar model}. As trade languages function as bridges across local languages and are widely spoken by local communities, according to the multipolar model, they can be well adapted as pivots for describing and interconnecting vernaculars. 

Again, we welcome Bird's idea of creating a multipolar model as the overarching system architecture for the development of linguistic knowledge; however, we propose to extend and to adapt his approach. Our difference lies primarily in the focus we are targeting: Bird and Schwarz put their focus on the question how to best work with small and disempowered indigenous communities (e.g., ~Australian Natives, Native Americans). While we agree that this focus is important, we add another perspective that centers around the self-critical analysis of the technologies that are currently being used, specifically their (often unintentional) by-design biases and the methodologies behind these biases. Therefore, we propose an additional focus on addressing not only linguistic, but also techno-linguistic bias or diversity, respectively. 

\section{Conclusion}
In this paper, we have shown that simply bridging the linguistic divide in language technology by extending existing systems does not serve the goal of promoting meaningful diversity, which we conceptually distinguish from diversity achieved through the generalization of technology. While generalized diversity is scalable but neglects the more subtle differences between languages, the former is about people's different social practices, their different worldviews, and the different situated knowledges embedded in different languages. We make these differences manifest by referring to lexical gaps that are emerging in recent language technologies and that relate to relevant domains of life such as kinship relations, educational systems, time, or food. Given these examples, we argue that extending existing language technologies to additional languages in this way can be even detrimental rather than beneficial to meaningful diversity.

Moreover, we point out that the problems of centralized development of language technologies go back deep into the colonial past, which is still potent today. Given these circumstances, there is an urgent need to be aware not only of the linguistic bias in language technologies that relates to the languages represented, but also to pay due attention to the techno-linguistic bias that relates to the design of the technologies. This is important because, as we argue, technological expansion on a basis such as we currently observe is not only detrimental to meaningful diversity, but actually contributes to a form of injustice that we identify as epistemic injustice. This form of injustice is not aimed at the distribution of epistemic goods, which is indeed officially encouraged by recent efforts to expand multilingual language technologies. Rather, this form of injustice is about the lack of recognition of certain forms of knowledge, modes of expression, and social realities that are evident in the lexical gaps we have identified. This form of injustice is not only problematic in itself, but also troubling in that it can be understood as an extension of colonial domination.

In light of these criticisms, we conclude that any efforts to extend existing language technologies that are not based on a rigorous approach to co-creation with the language communities in question should be fundamentally re-framed. By rigorous, we attribute approaches based on a critical stance toward the privileges of whiteness that avoids any kind of white \textit{savoir-faire} and instead conceives of the process as an opportunity for mutual learning in which neither party is superior to the other. Further empirical research, beyond this more overview-type study, is needed to better understand how techno-linguistic biases play out on the ground: in different circumstances and for different communities, as well as what specific sociopolitical consequences result for different groups of people. Such findings are important as they can shed light on key issues that need to be accounted for in future attempts to develop diversity-centered language technology.



\backmatter

\bibliography{sample-base}

\end{document}